\documentclass[twocolumn]{aa}
\usepackage{longtable}
\usepackage{graphics}

\def\deg{$^\circ$}
\def\arcsec{$''$}
\def\arcmin{$'$}
\def\hour{$^{\rm h}$}
\def\min{$^{\rm m}$}

\def\ha{H$\alpha$}
\def\nii{\mbox{[$\rm N \;{\scriptsize II}$]}}
\def\hi{\mbox{$\rm H \;{\scriptsize I}$}}
\def\hii{\mbox{$\rm H \;{\scriptsize II}$}}
\def\kms{$\rm km\;s^{-1}$}

\begin{document}

\thesaurus{     03(11.05.2; 
	   11.09.1 NGC 6221; 
	            11.09.2; 
		    11.11.1; 
		    11.19.2)}

\title{
Mixed Early and Late-Type Properties in the Bar of NGC~6221: 
Evidence for Evolution along the Hubble Sequence?\thanks{Based on observations 
collected at the European Southern Observatory, La Silla (Chile).} }

\author{J.C.~Vega Beltr\'an     \inst{1},
        W.W.~Zeilinger          \inst{2},
        P.~Amico                \inst{3},
        M.~Schultheis           \inst{2},
        E.M.~Corsini		\inst{4},
	J.G.~Funes, S.J.	\inst{5},
        J.~Beckman              \inst{6},
        F.~Bertola              \inst{5}
}

\offprints{Juan Carlos Vega Beltr\'an}
\mail{jvega@astrpd.pd.astro.it, jvega@ll.iac.es}

\institute{
Telescopio Nazionale Galileo, Osservatorio Astronomico di Padova, Padova, 
  Italy \and
Institut f\"ur Astronomie, Universit\"at Wien, Wien, Austria            \and
European Southern Observatory, Garching bei M{\"u}nchen, Germany        \and 
Osservatorio Astrofisico di Asiago, Dipartimento di Astronomia, Universit{\`a} 
di Padova, Asiago, Italy						\and
Dipartimento di Astronomia, Universit{\`a} di Padova, Padova, Italy     \and
Instituto de Astrof\'\i sica de Canarias, La Laguna, Spain 
}

\date{Received..................; accepted...................}
 
\titlerunning{Mixed Early and Late-Type Properties in the Bar of NGC~6221}
\authorrunning{Vega et al.}

\maketitle

\begin{abstract}

Rotation curves and velocity dispersion profiles are presented for both the
stellar and gaseous components along five different position angles
(P.A.=$5^{\circ}$, $50^{\circ}$, $95^{\circ}$, $125^{\circ}$, and $155^{\circ}$)
of the nearby barred spiral NGC~6221.
The observed kinematics extends out to about $80''$ from the nucleus.
Narrow and broad-band imaging is also presented.
The radial profiles of  the fluxes ratio \nii\ ($\lambda$ 6583.4 \AA )/\ha\ 
reveal the presence of a ring-like structure of ionized gas, with a radius 
of about $9''$ and a deprojected circular velocity of about 280 \kms .
The analysis of the dynamics of the bar indicates this ring
is related to the presence of an inner Lindblad resonance (ILR) at 1.3 kpc.  
NGC~6221 is found to exhibit intermediate properties between those
of the early-type barred galaxies: the presence of a gaseous ring at an ILR, 
the bar edge located between the ILR's and the corotation radius beyond 
the steep rising portion of the rotation curve, the dust-lane pattern,
and those of the late-type galaxies: an almost exponential surface brightness 
profile, the presence of \ha\ regions along all the bar, the spiral-arm pattern.
It is consistent with scenarios of bar-induced evolution from later
to earlier-type galaxies\footnote{Tables 3, 4, 5, 6, 7 and 8 are only
available in electronic form at the CDS via anonymous ftp to
cdsarc.u-strasbg.fr (130.79.128.5) or via
http://cdsweb.u-strasbg.fr/Abstract.html}.

 
\end{abstract}

\section{Introduction}

\object{NGC~6221} is a nearby spiral classified as Sbc(s) by Sandage \& 
Tammann (1981) and as SBc(s) by de Vaucouleurs et al. (1991). 
In The Carnagie Atlas of Galaxies (Panel~189)
Sandage and Bedke (1994) describe its morphology as semichaotic with two
symmetric heavy dust lanes starting from the nucleus and threading through 
the middle of the opposite thick arms, which begin at the centre.
They also stated that although the galaxy is not strongly barred its dust 
pattern is similar to that of prototype SBb galaxies, characterized by 
two straight dust lanes. 
An overview of the optical properties of the galaxy is given in Table~1.

NGC~6221 forms an apparent physical pair with the late-type spiral
\object{NGC~6215}. They have a systemic velocity of $1465\pm10$ \kms\ 
(this paper) and $1521\pm43$ \kms\ (RC3) respectively. Their angular distance 
is about 26$'$ (RC3) corresponding to a projected linear separation of about 
227 kpc at a mean distance of 30 Mpc ($H_0 = 50\;\rm km\;s^{-1}\;Mpc^{-1}$)
The galaxy is also possibly interacting with two newly discovered
low-surface brightness galaxies nearby (Koribalski 1996a). 

The dynamics of the ionized gas component in this galaxy has been studied by
Pence \& Blackman  in 1984 (hereafter P\&B). They measured a conspicuous
S-shaped pattern in the ionized gas velocity field, showing that large 
velocity gradients occur at the position of the dust lanes in the bar. 
They have been interpreted as the signature of shock fronts of the gas
which reverses its motion from outward to inward as it passes through them.

Philipps (1979) reported that the nuclear spectrum of NGC~6221 exhibits
signatures of emission originating from \hii\ regions and also a weak
Seyfert~2 component.
He concluded however that the emission line spectrum can be interpreted as 
gas being ionized by hot young stars without the need to introduce a 
non-thermal component in the galaxy nucleus. The analysis of the spatial 
distribution of the ionized gas  (Durret \& Bergeron 1987) confirmed this 
picture. Some dozens of \hii\ regions were identified in the disk region 
extending as far as 9~kpc from the galaxy nucleus. Although the emission in
the nucleus appears to be extended ($\sim 1 \; {\rm kpc}^2$), the authors
conclude that the main ionizing sources are OB stars and that UV
and X-ray radiation from the nuclear source might not penetrate the
heavily obscured nuclear region. Dottori et al. (1996) also recently
suggested that the nucleus of NGC~6221 may harbour sources of type \hii\ 
and also a Seyfert~2 source. 
 
In this paper we study the kinematics of the ionized gas and the stellar
components in NGC~6221. We present the velocity curves and the velocity 
dispersion profiles of gas and stars obtained along five different position
angles by means of long-slit spectroscopy.

\begin{figure}
\resizebox{\hsize}{!}{\includegraphics{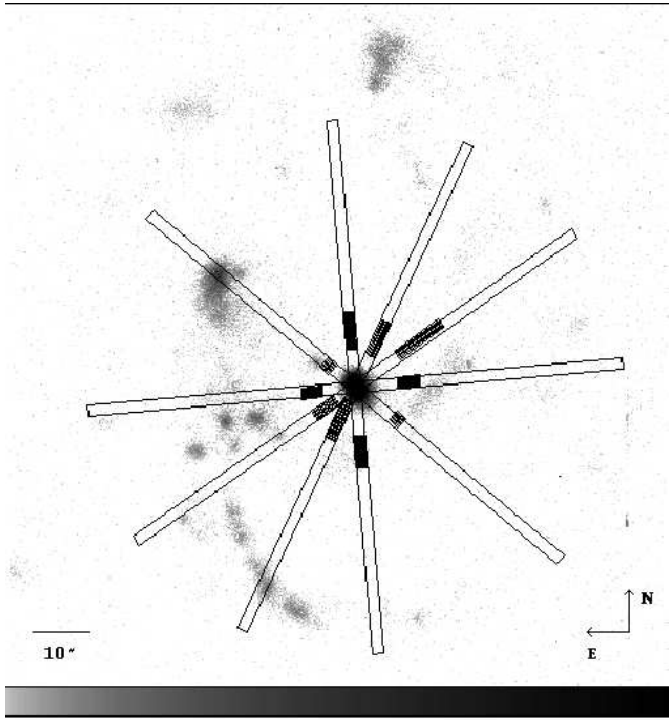}}
\caption{\ha$+$\nii\ emission lines image of NGC~6221.
The slit positions (P.A.=$5^{\circ}$, $50^{\circ}$,
$95^{\circ}$, $125^{\circ}$, and $155^{\circ}$) of the obtained spectra 
are plotted. 
The {\it black bars\/} show the position and the size of the regions 
(along the different slit positions) where the \nii/\ha\ 
fluxes ratio is greater than 0.5. The image orientation and scale are 
indicated 
}
\label{fig:halpha}
\end{figure}

\section{Observations and data reduction}

\subsection{Narrow and Broad-Band Imaging}

Two narrow-band images of NGC~6221 were extracted from the ESO NTT archive.
They were taken on the night of May 6, 1994 with the NTT Telescope at La Silla.
The No.~25 Tektronix TK1024M CCD with $1024\times1024$ pixels was used as
detector in combination with SUSI. Each 24$\mu$m$\times24\mu$m image pixel 
corresponds to 0.13$''\times0.13''$. The CCD gain and the readout noise were 
determined to be 3.4 e$^{-}$ per ADU and 5.9 e$^{-}$, respectively.  

The emission-band image was obtained with the interference ESO filter No.~692
isolating the spectral region containing the redshifted \ha\ and \nii\ 
($\lambda$ 6583.4 \AA) emission lines. The continuum-band image was taken 
with the interference ESO filter No.~696 isolating an emission-free spectral 
region. The exposure time was 20 minutes for each filter.  The seeing was 
about $1.7''$

\begin{figure}
\resizebox{\hsize}{!}{\includegraphics{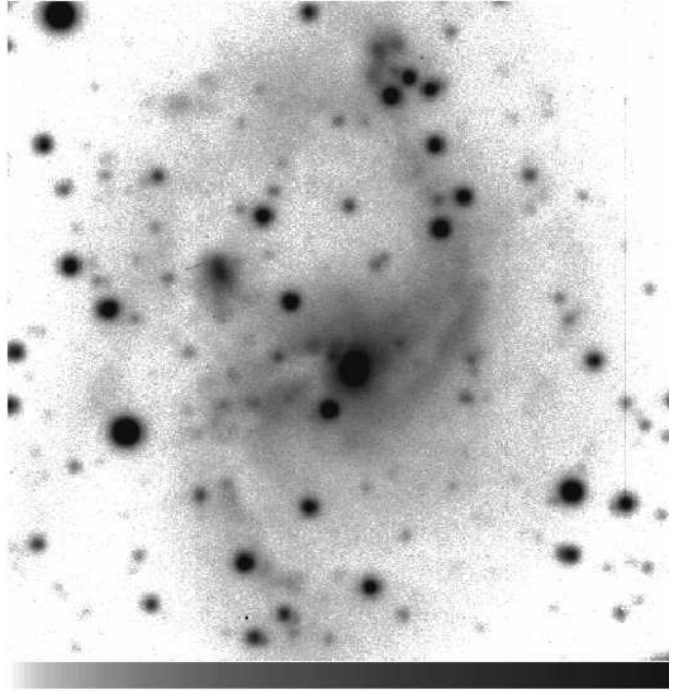}}
\caption{Stellar continuum image of NGC~6221. The image orientation and
scale are the same as in Fig.~1  
}
\label{fig:continuum}
\end{figure}

\begin{table}[t]
\caption{Optical properties of NGC~6221}
\begin{flushleft}
\begin{tabular}{lc}
\hline
\noalign{\smallskip} 
parameter & value \\
\noalign{\smallskip} 
\hline
\noalign{\smallskip} 
Name                                 &  NGC~6221			 \\
Morphological type                   &  Sbc(s)$^{\rm a}$; .SBS5$^{\rm b}$ 	\\ 
Position (equinox 2000.0)$^{\rm b}$  & \\
\hspace{.2truecm} right ascension $\alpha$  &  16\hour$\;$52\min$\;46\fs8$ \\
\hspace{.2truecm} declination     $\delta$  &  $-59$\deg$\;$12\arcmin$\;$59\arcsec \\
Heliocentric systemic velocity $cz$$^{\rm c}$  & $1465\pm10$ \kms \\
Position angle P.A$^{\rm b}$               &  5\deg \\
Isophotal diameters D$_{25}\times$d$_{25}$$^{\rm b}$ &              
 $3\farcm55\times2\farcm45$ \\
Inclination $i^{\rm d}$                   &  44\deg  \\
Total corrected $B$ magnitude $B_T^0$$^{\rm b}$     &  $10.66$ mag \\
\noalign{\smallskip} 
\hline
\end{tabular}
\begin{list}{}{}
\item[$^{\rm a}$] from Sandage \& Tamman (1981)
\item[$^{\rm b}$] from de Vaucouleurs et al. (1991)
\item[$^{\rm c}$] from this paper 
\item[$^{\rm d}$] from log R$_{25}$ of de Vaucouleurs et al. (1991)
\end{list}
\end{flushleft}
\end{table}

Standard reduction of the two images was performed using the original 
available data. The two images were aligned using field stars as reference 
(with an accuracy better than 0.2 pixel). In each image the sky level was 
estimated in areas of the frames unaffected by the galaxy light and then 
subtracted. The continuum-band image was subtracted from the emission-band image
after an appropriate intensity scaling, allowing for transmission differences 
between the filters. No attempt was made to calibrate in flux the resulting
pure-emission image of NGC~6221. 

Two broad-band images of NGC~6221 in the $I$ and the $J$ band respectively
were extracted from the DENIS data archive (Epchtein et al. 1997). 
Their reduction was routine.

The reduction of the broad and narrow-band images and the surface
photometry of the NTT continuum and DENIS $I$ and $J$-band frames was 
performed by means of standard procedures within the ESO-MIDAS software
package. 
 
\subsection{Long-Slit Spectroscopy}

The spectroscopic observations of NGC~6221  were carried out at the ESO 1.52~m
Spectroscopic Telescope at La Silla using the Cassegrain Boller \& Chivens
Spectrograph in the nights of April 30 and May 2, 1992.
The No.~26 1200 $\rm grooves\;mm^{-1}$ grating blazed at 5730 \AA, was 
used in the first order in combination with a 2.5$''\times2.1'$ slit.  
The No.~24 FA2048L CCD with 2048$\times$2048 pixels
was adopted as detector. It yielded a wavelength coverage of $\sim2000$ \AA\ 
between about 4900 and about 6900 \AA\ with a reciprocal dispersion of 65.1 
$\rm \AA\;mm^{-1}$. The instrumental resolution was determined measuring 
the FWHM of the emission lines of the comparison spectra. We checked that 
such measured FWHM's do not depend on wavelength and we found a mean value 
of FWHM = 2.35 \AA\ (i.e. $\sigma = 1.0$ \AA\ which corresponds in the range
of the observed emission lines to $\sim50$ \kms). Each 15$\mu$m$\times15\mu$m
spectrum pixel corresponds to 0.98\AA $\times 0.81''$.

The long-slit spectra of NGC~6221 were obtained at different position angles 
in order to map the velocity field. The observing log is given in Table~2. 
The seeing value was about $1.2''$. Each object spectrum was bracketed by
two helium-argon calibration spectra. The stars HR~3431 (K4 III),  HR~5601 
(K0.5 III), HR~6318 (K4 III) and HR~7597 (K0 III) were observed as velocity 
templates. The standard spectral reduction was performed by using the 
ESO-MIDAS package. All the spectra were bias subtracted, flat-field
corrected by quartz lamp exposures and wavelength calibrated by fitting the 
position of the comparison lines with cubic polynomials.
Pixels affected by cosmic ray events were identified and then corrected.
The contribution of night sky was determined from the edge regions 
(not contaminated by galaxy light) of each spectrum and then subtracted.

\begin{table}
\begin{flushleft}
\caption{Observing log for NGC~6221}
\begin{tabular}[t]{crcl}\hline \noalign{\smallskip}
  \multicolumn{1}{c}{Date}       & \multicolumn{1}{c}{PA}
& \multicolumn{1}{c}{exp.~time}  
& \multicolumn{1}{c}{comments} \\
  \multicolumn{1}{c}{ }          & \multicolumn{1}{c}{[$^\circ$]} 
& \multicolumn{1}{c}{[$^s$]}      
& \multicolumn{1}{c}{ } \\
\noalign{\smallskip}\hline \noalign{\smallskip} 
30$-$Apr$-$1992 &   5 & 3600  & major axis \\
02$-$May$-$1992 &  50 & 3300  & \\
02$-$May$-$1992 &  95 & 3600  & minor axis \\
02$-$May$-$1992 & 125 & 3300  & bar major axis\\
30$-$Apr$-$1992 & 155 & 3600  & \\
\noalign{\smallskip} \hline
\end{tabular}
\end{flushleft}
\end{table}

The stellar velocities and velocity dispersions were measured from the 
absorption lines in the wavelength range between about 4900 \AA\ and 6200 \AA\
using the Fourier Correlation Quotient technique (Bender 1990) as applied by 
Bender et al. (1994). The measured radial velocities were corrected to the 
heliocentric frame of reference. The heliocentric correction was 
$\Delta v=+15.6$ \kms\ and $+14.9$ \kms\ for April, 30 and May, 2 respectively.

The stellar kinematics measured along the different position angles is
given in Table~3. The table provides  the position angle (Col.~1),
the radial distance from the galaxy centre in arcsec (Col.~2), the
observed heliocentric velocity with its respective error (Cols.~3,
and 4),  and the velocity dispersion  with errors (Cols.~5 and 6) in
\kms, and the Gauss-Hermite coefficients wtih errors $h_3$ (Cols.~7 and 8) and
$h_4$ (Cols.~9 and 10).

The gas velocities and velocity dispersions were measured by means of the
MIDAS package ALICE from the \ha\ and \nii\ ($\lambda$ 6583.4 \AA)
emission lines. The position, the FWHM and the uncalibrated flux of each 
emission line  were individually determined by interactively fitting a
Gaussian plus a second order polynomial to the emission and its surrounding
continuum. The wavelength of the Gaussian centre was converted to the velocity
$cz$, and then the heliocentric correction was applied. The velocity errors 
were derived as in Bertola et al. (1996). The Gaussian FWHM was corrected for
the instrumental FWHM and then converted to the velocity dispersion 
$c\sigma/\lambda_0$. The velocities and the velocity dispersions derived 
from the two emission lines agree within 5\%.
At each radius the weighted mean of the velocities and the average of the 
velocity dispersions were taken as the velocity and velocity dispersion 
of the NGC~6221 ionized gas respectively.

The ionized gas kinematics measured along the different position angles
is given in the five Tables~4--8. Each table provides the radial
distance from the galaxy centre in arcsec (Col.~1), the \ha\ observed
heliocentric velocity with its respective error (Cols.~2 and 3) and
velocity dispersion (Col.~4) in \kms, the \nii\ observed heliocentric
velocity with error (Cols.~5 and 6) and velocity dispersion (Col.~7) in
\kms, the \nii/\ha\ flux ratio (Col.~8), and the averaged ionized gas
heliocentric velocity with error (Cols.~9 and 10) and velocity
dispersion with error (Cols.~11 and 12) in \kms.

\section{Results}

\subsection{The Surface Photometry}

The spatial distribution of the \ha$+$\nii\ emission is displayed in Fig.~1
together with the slit positions of the long-slit spectra. Most of the 
emission originates from knots in the spiral arm regions. These knots are 
the typical signatures of the presence of \hii\ regions. The emission is also
found to originate close to the dust lanes.  The complex dust lane structures in
the area of the bulge and bar are evident in the continuum image of NGC~6221 
(Fig.~2). We estimate a bar projected radius of about $20''-25''$ in agreement
with Elmegreen \& Elmegreen (1985).

\begin{figure}
\resizebox{\hsize}{!}{\includegraphics{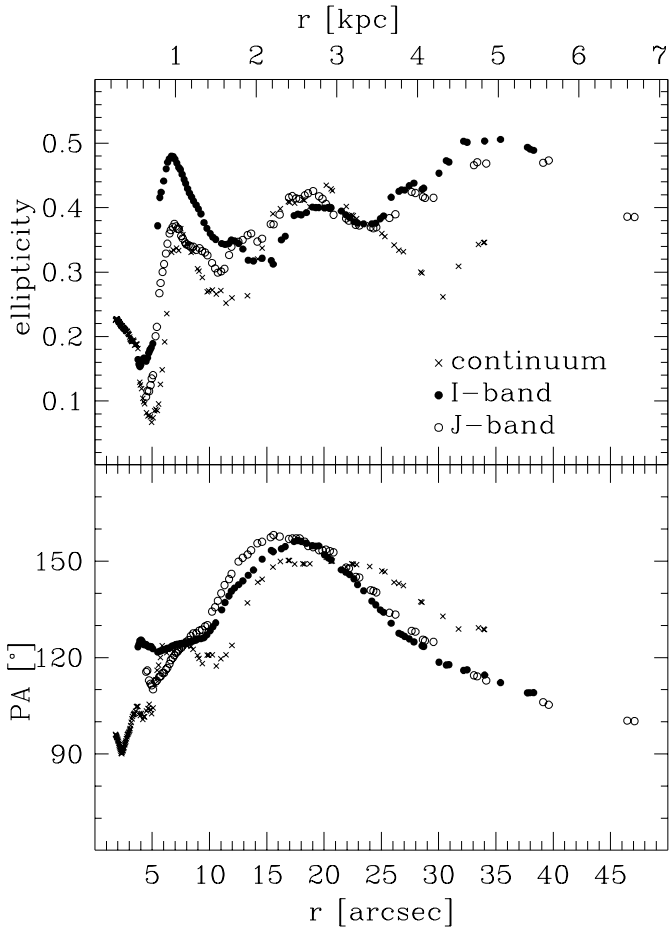}}
\caption{NGC~6221 ellipticity and position angle radial profiles obtained
from the ellipse fitting to the NTT continuum-band ({\it crosses\/}),
DENIS {\it I\/}-band ({\it filled diamonds\/}), and  DENIS {\it J\/}-band 
({\it open circles\/}) images 
}
\end{figure}

From the isophotal analysis  of the continuum image and the two DENIS images 
in the bands I and J, the position of the bar was determined. 
The maximum in ellipticity and the plateau of the position angle profile
around P.A.$=125^\circ$ (Fig.~3) were used as indicators of the bar component. 
This position angle value is also consistent with the orientation of the 
dust lanes in Fig.~2. P\&B derive for NGC~6221 a position angle for the 
line-of-nodes of $118^\circ \pm 5^\circ$. They also stated that the major axis
of the bar is exactly perpendicular to the line-of-nodes. We believe they 
confused the position angles of the major axis of galaxy with that of the bar
as can be seen by Fig.~2 in this paper and by Plates~2 and 3 in their
paper and from our velocity field.
The uncalibrated surface brightness profile of the bar shows an exponential 
decrease, although heavily influenced by the presence of dust lanes.

\begin{figure}[h]
\resizebox{\hsize}{!}{\includegraphics{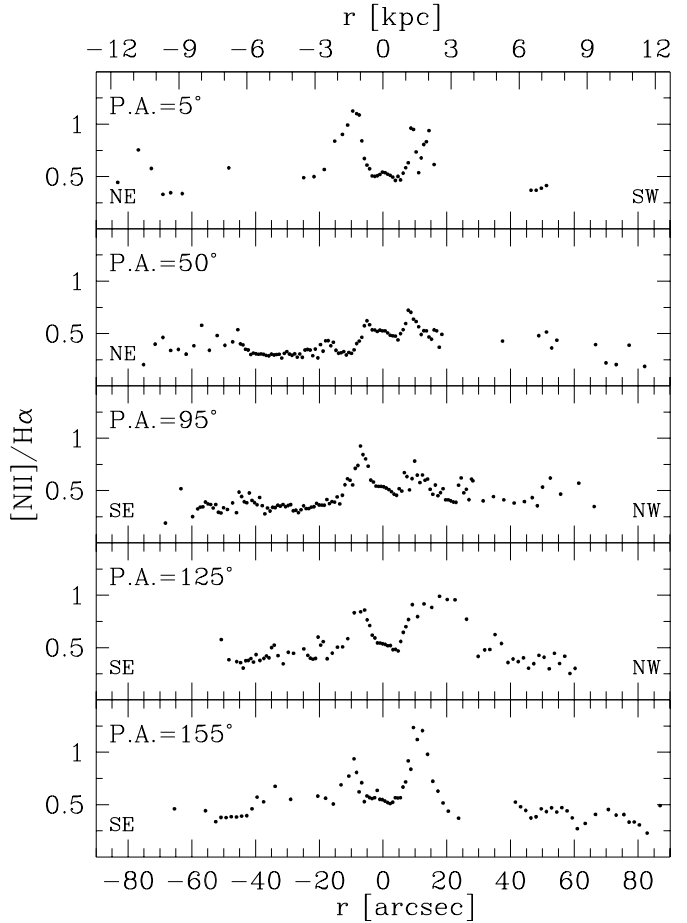}}
\caption{[N~II]($\lambda$ 6583.4 \AA )/\ha\ 
emission line fluxes ratios of
NGC~6221 as a function of radius for the
position angles 5$^\circ$ (major axis), 50$^\circ$, 95$^\circ$
(minor axis), 125$^\circ$ (bar major axis) and 155$^\circ$
}
\end{figure}

\subsection{The Radial Profiles of the \nii/\ha\ Ratio}

\begin{figure*}[ht]
\vspace*{11cm}
\includegraphics{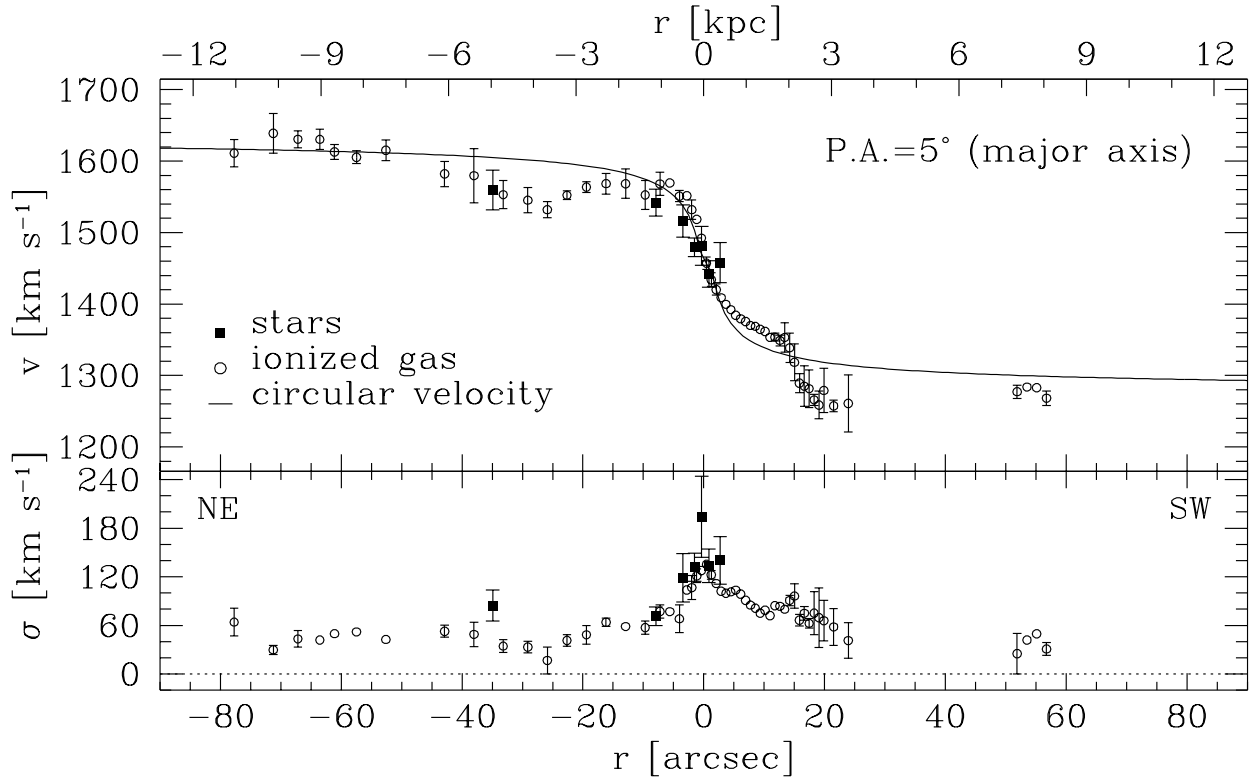}
\vspace*{.25cm}
\caption[]{The kinematics of the ionized gas ({\em open circles\/}) and
stars ({\em filled squares\/}) measures along NGC~6221 major-axis 
(P.A.=5$^{\circ}$). 
The observed heliocentric velocity curves and the velocity dispersion profiles
are shown in the top and in the bottom panel respectively. 
The errorbars of the gas velocities and velocity dispersions smaller 
than symbols are not plotted. 
The {\it solid line\/} represents the model circular velocity, as explained 
in the text 
}
\label{fig:p05}
\end{figure*}

The radial profiles of the fluxes ratio \nii\ ($\lambda$ 6583.4 \AA )/\ha\ 
were calculated for all position angles and are presented in Fig.~4. 
The value of the ratio  (\nii/\ha $\sim0.3$) indicates that the emission
is mainly from \hii\ regions, even in the bulge. The central regions where 
$0.5<$\nii/\ha$\leq1$ are marked as black bars along each slit position 
in Fig.~1. They constitute a ring-like structure with a radius of about $9''$
($\sim1$ kpc for the assumed distance), which is  is located at the distance
from the centre corresponding to the position of a ILR due to the presence 
of the bar (see \S~3.4 for the discussion).  The presence of this nuclear 
gaseous ring is consistent with the  \hi\ observations of Koribalsky (1996b).

\subsection{The Ionized Gas and Stellar Kinematics}

The measured ionized gas and stellar kinematics of the galaxy NGC~6221 are presented for 
the observed position angles in Figs.~5--9. 

In all the spectra we found an offset of $4''-5''$ between the stellar 
kinematical centre (obtained from the position of the maximum intensity of
continuum)  and the gaseous kinematical centre (estimated from the maximum 
emission of the \ha\ and \nii\ lines).  

The position of the gaseous kinematical centre coincides with that estimated
by P\&B. For to the systemic velocity we measured $V_{\rm sys} = 1465\pm10$  
while P\&B obtained a slightly larger value of $1475\pm5$. If we use the 
emission lines data as reference for the centre, we find for all intermediate
position angles that the folded velocity and velocity dispersion curves of 
ionized gas and stars present strong asymmetries. On the other hand, using 
the absorption lines data as reference for the centre the asymmetries of 
the folded kinematical profiles are minimized. In the following the absorption
line data are therefore used to give the true centre for both ionized gas
and stars.

In order to bring out the asymmetries in the gas velocity curves we compare 
them with the empirically derived curve of circular velocity projected onto 
the different position angles. We adopted the equation of the rotation curve 
of galaxy represented as a sequence of flattened spheroids of Brandt (1960)  
as applied by Bettoni \& Galletta (1997) for their sample of barred galaxies.
The parameters of the curve were found by the best fit  to the major axis NE 
side of ionized gas rotation curve (which has the more regular pattern) 
assuming a galaxy inclination of $44^\circ$ (RC3). The resulting curve has 
been folded for the major axis SW side. Then it has been projected onto the
remaining the position angles assuming that the mean motions were symmetric 
with respect to the galaxy rotation axis. These projected curves 
(see Figs.~5-9) allow us to estimate when ionized gas and stars deviate from 
circular motions.

\subsubsection{The Major Axis (P.A.$=5^\circ$)}

The kinematics of the ionized gas and starsmeasured along the NGC~6221 optical 
major axis is shown in Fig.~5.

\begin{figure*}[ht]
\vspace*{11cm}
\includegraphics{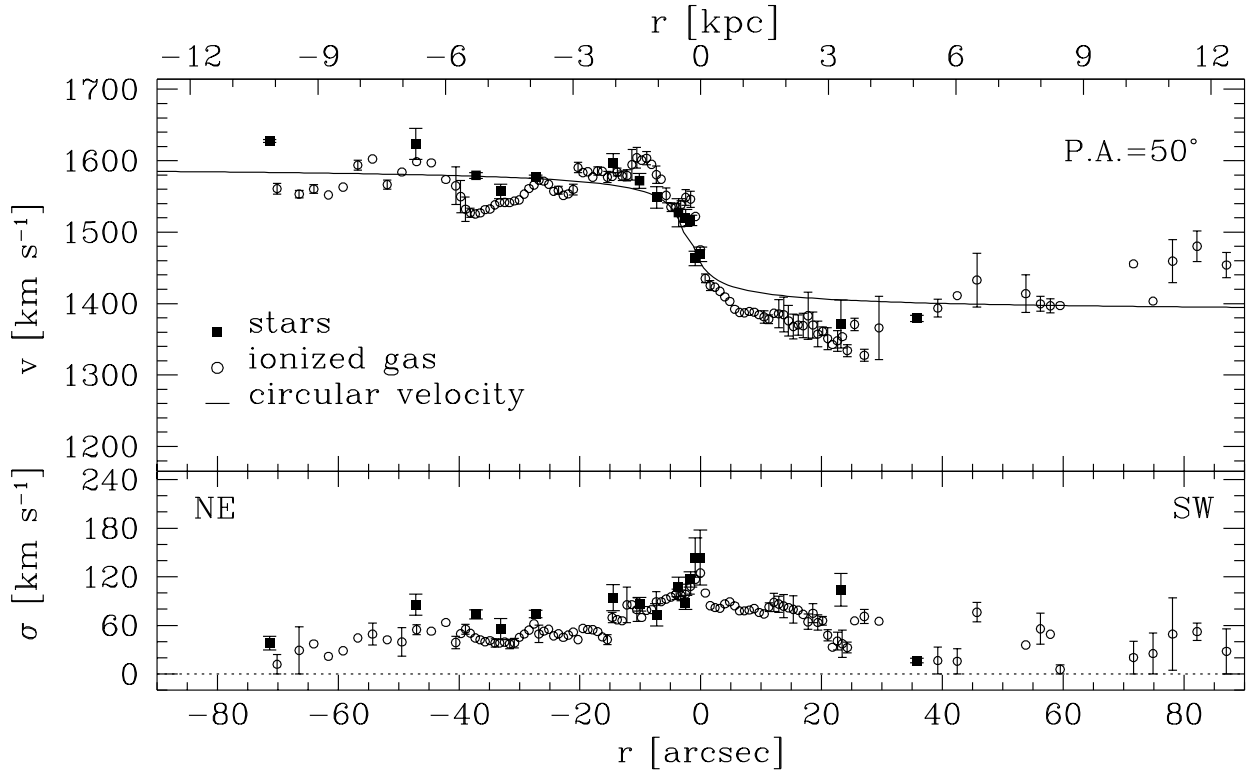}
\vspace*{-.5cm}
\caption[]{Same as Fig.~5 for P.A.=50$^{\circ}$
}
\label{fig:p50}
\end{figure*}

\begin{figure*}[ht]
\vspace*{11cm}
\includegraphics{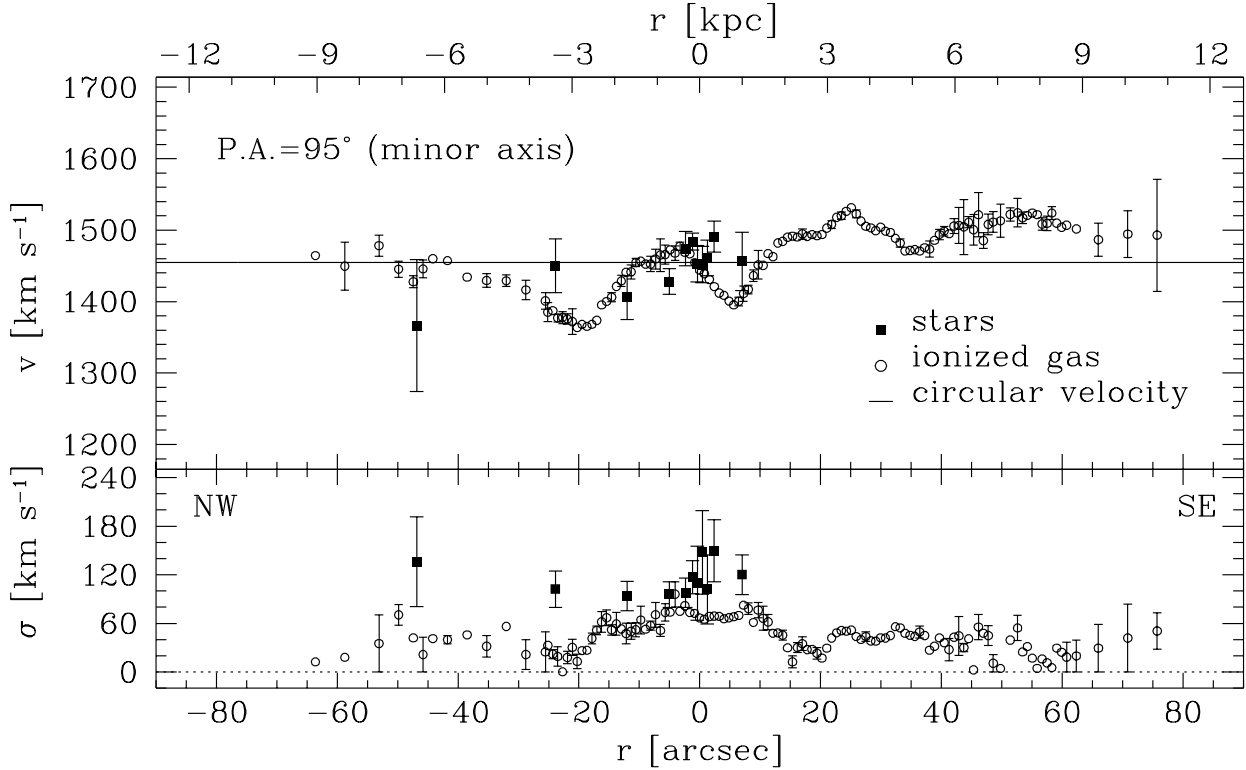}
\vspace*{-.5cm}
\caption[]{  Same as Fig.~5 for P.A.=95$^{\circ}$ (minor axis)
}
\label{fig:p95}
\end{figure*}

\begin{figure*}[ht]
\vspace*{11cm}
\includegraphics{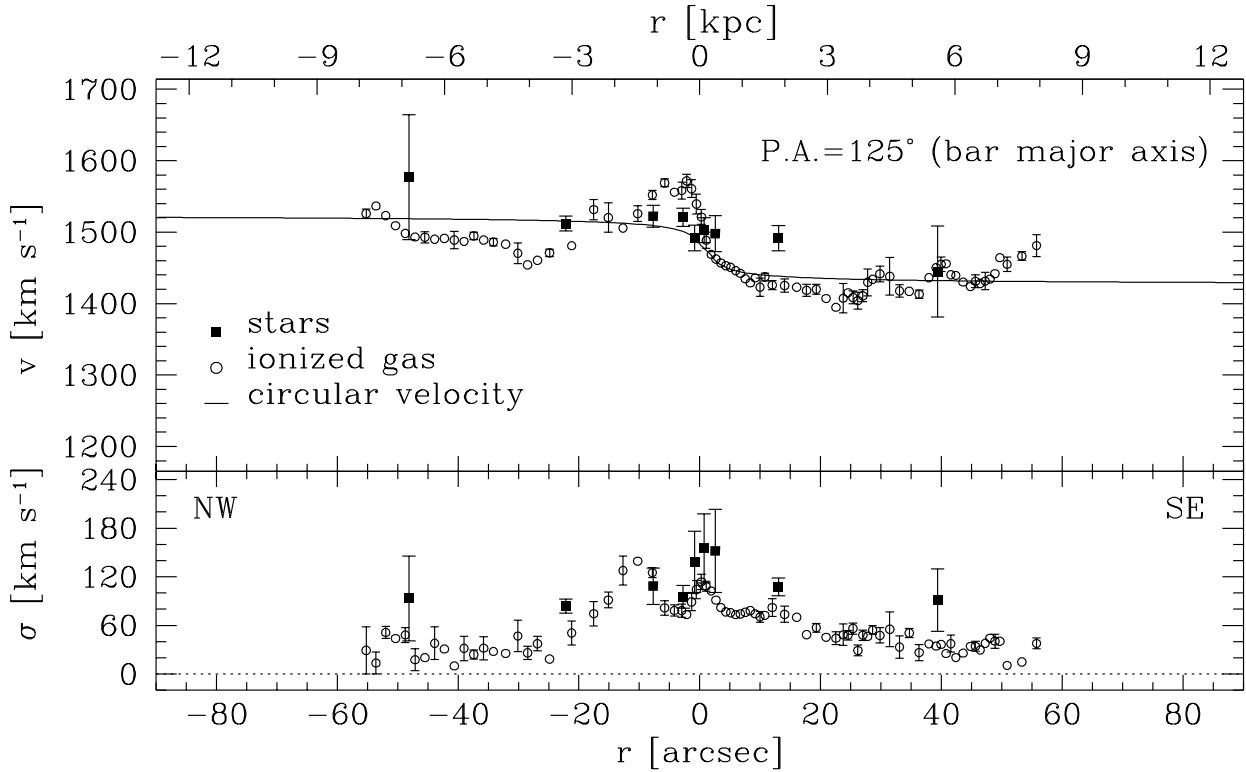}
\vspace*{-.5cm}
\caption[]{Same as Fig.~5 for P.A.=125$^{\circ}$ (bar major axis)}
\label{fig:p125}
\end{figure*}

\begin{figure*}[ht]
\vspace*{11cm}
\includegraphics{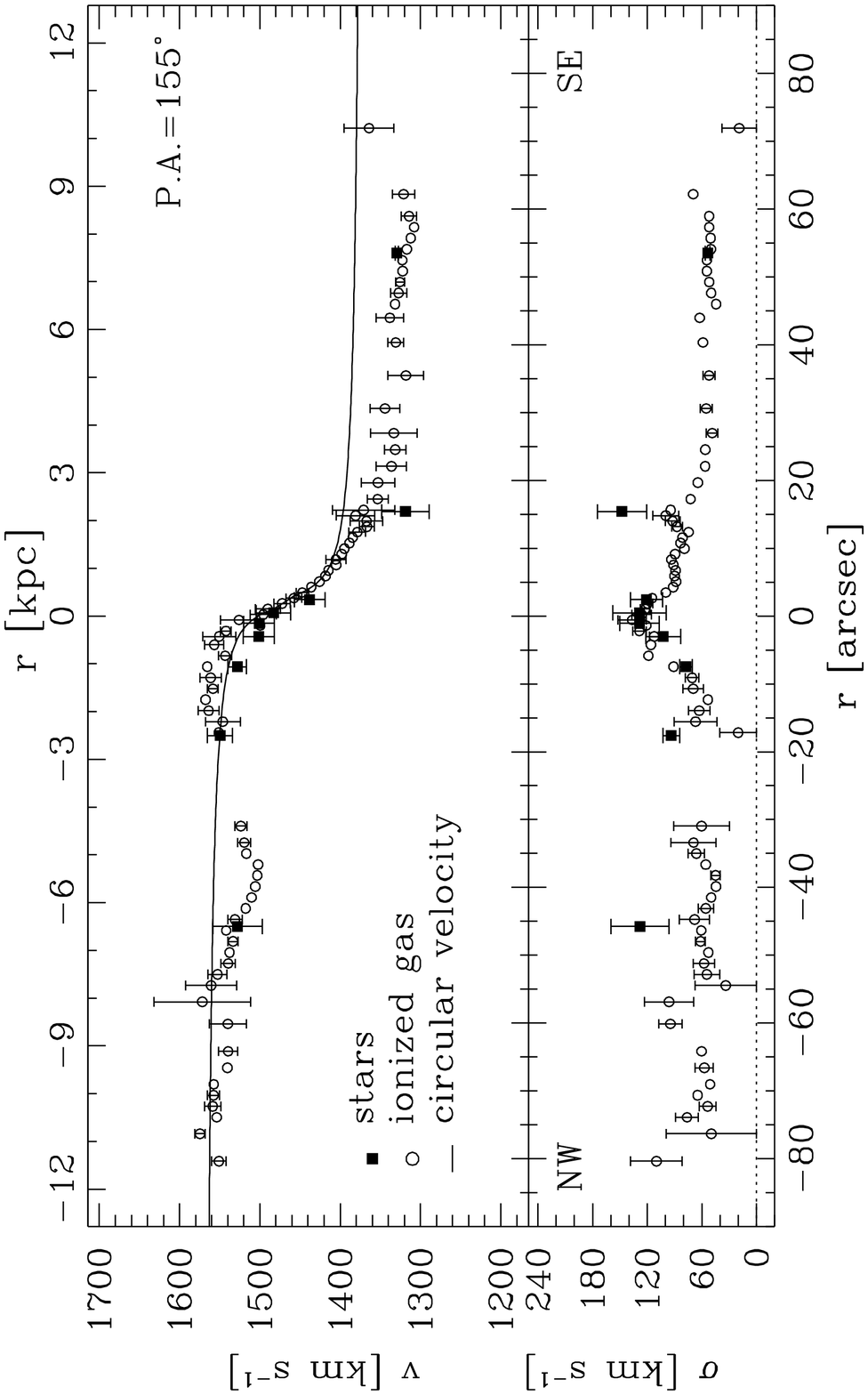}
\vspace*{-.5cm}
\caption[]{Same as Fig.~5 for P.A.=155$^{\circ}$}
\label{fig:p155}
\end{figure*}

The ionized gas kinematics extends to about $80''$ ($\sim11$ kpc) on the 
receding SE side and to about $55''$ ($\sim8$ kpc) on the approaching SW side.
We found that our gas rotational velocity data agrees within the errors with
the measurements obtained by P\&B along their spectrum G (P.A.$=14^\circ$) 
and deprojected to our position angle. 
We obtain more detailed measurements in the central region where the velocity
gradient is strong ($\Delta v_{\rm gas} = 350$ \kms\ for $\Delta r=20''$).
The velocity dispersion of the gas show within the bulge a peaked trend for
$|r|<5$ with a maximum value of about 130 \kms. In the outer region dominated
by the disk component the velocity dispersion is found to be constant at a 
value of about 50 \kms. Similar peaked velocity dispersion profiles are also
found in bulges of elliptical (e.g. Zeilinger et al. 1996) and lenticular 
galaxies (e.g. Bertola et al. 1995). For $15''<|r|<25''$ the gas velocity 
curve is very asymmetric as shown by the comparison with the empirical 
circular velocity curve. In this radial range the gas velocity along the 
receding side is about 20 \kms\ lower then the circular velocity. It is about
60 \kms\ higher along the approaching side. An abrupt change in the slope of
the gas velocity curve and a local maximum in the gas velocity dispersion is
also observed at about $17''$. No emission is detected for $25''<r<50''$. 

The stellar kinematics data extend from about $35''$ ($\sim5$ kpc) on the 
SE side to near the centre of the galaxy. The velocities of the stars follow
closely the trend of the velocities of the gas.  The stellar velocity 
dispersion profile is peaked with a central value of about 200 \kms. At larger
radii it drops off to a value of about 70 \kms\ along the NE side. For the 
SW side it was not possible to measure the stellar kinematics due to the 
poor $S/N$ ratio of the absorption lines.

The stellar kinematics of the bulge is found to be consistent with a hot 
stellar system ($v/\sigma\simeq0.3$ at $\epsilon\simeq0.2$). The fact that 
the velocity dispersion of the gaseous component is  of the same order as
that of the stellar one indicates a significant contribution of random motions
to the dynamical support of the gas as dicussed by Bertola et al. (1995) for 
a sample of S0 galaxies.

\subsubsection{The Intermediate Axis at P.A.$=50^\circ$}

The kinematics of ionized gas and stars measured along the NGC~6221 
P.A.$=50^\circ$ is shown in Fig.~6.

The gas kinematics is observed to about $70''$ ($\sim10$ kpc) on the NE side
and to about $85''$ ($\sim12$ kpc) on the other side. For this intermediate 
axis our gas velocity data agrees within the errors with those of P\&B 
(spectrum J). The central velocity gradient ($\Delta v_{\rm gas} = 300$ \kms\
for $\Delta r = 20''$) is lower than on the major axis. The gas velocity 
dispersion has a central maximum of about 120 \kms. It decreases slowly to 
about 40 \kms\ with the distance from the centre. 

The stellar velocity curve is close to the gaseous one. Except for 
$-10''<r<-3''$ the stellar velocity dispersion remains higher than the 
gaseous one of about 20 \kms . The stellar data are as extended as the gas
data in the NE side and limited to about $35''$ ($\sim5$ kpc). The SW side 
is characterized by regions without emission lines and with low $S/N$ 
absoption lines. This  produces the large errors in the measurement of the 
velocity of the gas  and makes impossible to determine the stellar kinematics.

\subsubsection{The Minor Axis (P.A.$=95^\circ$)}

The kinematics of ionized gas and stars measured along the NGC~6221 optical 
minor axis are shown in Fig.~7.

The gas kinematics is measured to about $65''$ ($\sim9$ kpc) and to about 
$75''$ ($\sim11$ kpc) on the NW and SE side respectively. The gas velocity 
measurements obtained by P\&B along their spectrum B  (P.A.$=104^\circ$) 
deprojected to P.A.$=95^\circ$ agree within the data errors with the 
velocities we found along this axis. If we assume an axisymmetric
potential for a galaxy we expect to find a zero velocity gradient along
the minor axis; nevertheless, here we observe a `wavy pattern' in the gas
 velocity curve at this
position angle. P\&B defined this behaviour as a `conspicuous S-shaped
pattern' on the isovelocity contours of the gas. For $|r|>30''$ the gas 
velocity follows the circular velocity only on the NW side. The folded velocity 
curve is asymmetric for $|r|>10''$, with a maximum velocity difference 
between the NW and SE side of about 60 \kms  . 

In spite of the fact that the minor axis differs by about $30^\circ$  from
the bar position angle,  the virtual absence of 
projected circular motions at this angle
makes it possible to study in great detail how the bar affects the dynamics
of gas and stars. The non-circular motions have an observed maximum amplitude
of about 100 \kms, which is found to be in agreement with the rather general
predictions  by Roberts et al. (1979), who estimate velocities in the 
range of 50 \kms\ to 150 \kms\ for bar-induced radial components of gas 
flow. 
From the analysis of Fig.~2 a  projected radial extent of $20''-25''$ 
is estimated for the bar component, and this is found to
be in agreement with the kinematical data. The gaseous velocity dispersion
profile in the region of the bar is found to be constant with a value of
about 70 \kms\, distinctly different from the two preceding 
postion angles (i.e. at $5^\circ$ and $50^\circ$) where
the profile was peaked at the centre. Outside the bar region the gas
velocity dispersion is lower than 50 \kms . 

From the above, we can explain the `wavy pattern' in the gas velocity curve
along the minor axis as follows. For the central $|r|<6''$ the velocity of
the gas shows a negative gradient because it follows the $x_1$ family orbits, 
a little further out the gradient is inverted on both sides of the velocity 
curve. This can be explained if we consider that beyond the position of the
outer ILR (located at a deprojected radius $r\sim9''$) the gas follow the
$x_2$ family orbits out to the corotation radius (at a deprojected radius 
$r\sim33''$). Outside the corotation the bar influence declines and the 
velocity of the gas tends to the systemic velocity, at least on the NW side. 

The star kinematics is measured up to a distance of about $45''$ ($\sim7$ kpc)
in the NW side and up to about $7''$ ($\sim1$ kpc) in the SE side.
The velocity  curve for the stars agrees within the error limits with the 
circular velocity. At this angle the velocity dispersion appears significantly
higher for the stars than for the gas. The stellar velocity dispersion is 
higher than 90 \kms\ at all radii peaking to about 150 \kms\ in the 
centre. 
For the SE side ($r>10''$) of the spectra it was not possible to obtain
measurements of the stellar kinematics.

\subsubsection{The Bar Major Axis (P.A.$=125^\circ$)}

The kinematics of ionized gas and stars measured along the NGC~6221 
P.A.$=50^\circ$ are shown in Fig.~8. From the photometrical data we deduced 
that this is the position angle of the major axis of NGC~6221 bar. 

The gas kinematics extends to about $55''$ ($\sim8$ kpc) from the nucleus
on either side. P\&B did not obtain gas velocity measures for this angle.  
The gas velocity curve is strongly asymmetric. It follows the predicted 
circular velocity only along the approching SE side. On the receding NW side
the observed gas velocity is lower ($-50''<r<-20''$), equal ($-20''<r<-10''$)
and higher ($-10''<r<0$) than the circular velocity. The observed maximum 
deviation from the empirical circular velocity is about 80 \kms\ at 
$r\sim-3''$. The gas velocity dispersion profile has an `M-shaped' appearance
due to the presence of two local maxima. It peaks at about 120 \kms\ in the
centre and at about 140 \kms\ at $r=-10''$. For the outer regions it decreases
and it becomes constant with a value lower than 40 \kms. 

The stellar kinematics is observed to about $50''$ ($\sim7$ kpc) on the NW 
side and to about $40''$ ($\sim6$ kpc) on the SE side. The stars rotate more
slowy than gas with a higher velocity dispersion at all radii. This has a 
central maximum of about 160 \kms, and remains higher than 80 \kms\ at larger 
radii.

\subsubsection{The Intermediate Axis at P.A.$=155^\circ$}

The kinematics of ionized gas and stars measured along the NGC~6221 
P.A.$=155^\circ$ are shown in Fig.~9.

The gas kinematics is measured up to $80''$ ($\sim11$ kpc) and up to $70''$ 
($\sim10$ kpc) in the approaching and in the receding side respectively.
P\&B do not present gas velocity data are at this position angle.
The gas velocity curve has the same appearance as that along the major axis,
with the exception of a lower velocity gradient. Along all the SE side the 
observed velocity differs from the circular value by about 60 \kms. 
The gas velocity dispersion profile has a central maximum of about 140 \kms\ 
decreasing to values lower than 60 \kms\ for $|r|>20''$.

The stellar kinematics is observed to about $50''$ ($\sim7$ kpc) on either 
side of the centre. The stars rotate with similar velocity to the gaseous 
component. They also show the same velocity dispersion profile as the gas.
The  $S/N$ ratio for the absorption lines is very low for the SE side where 
the stellar kinematics is measured only at about $50''$ from the centre.

\subsection{The Location of the Lindblad Resonances}

With the empirical curve of the circular velocity $V(R)$ derived in the 
section \S~3.3 we are able to calculate the angular velocity
$\Omega(R)=V(R)/R$ and the epicyclic frequency  
$k(R) = (4\Omega^2+R{\rm d}\,\Omega^2/\,{\rm d}r)^{0.5}$ where $R$ is
the distance from the centre deprojected on the galaxy plane.
In Fig.~10 the curves $\Omega$, $\Omega + k/2$ and $\Omega - k/2$ are plotted
in the inner $70''$ ($\sim10$ kpc) from the centre.

If we assume the observed ring of ionized gas to be at the outer ILR 
we derive the pattern speed of the bar $\Omega_{p}$
from the condition $\Omega_{p}=\Omega\,(R_{\rm OILR})-k\,(R_{\rm OILR})/2$.
We have $R_{\rm OILR} = 9''$ ($\sim1$ kpc) and we find for the bar of NGC~6221
that $\Omega_{p}=46$ \kms\ kpc$^{-1}$ taking in account all the uncertainties. 
For this value of $\Omega_{p}$ there is another intersection with the curve
$\Omega - k/2$ giving the location of the inner ILR.
The radius of the inner ILR is small ($R_{\rm IILR} = 3'' \sim0.4$ kpc),
and its precise location is uncertain as it is comparable to the seeing 
resolution of our observation.

Corotation is located at the radius $R_{\rm CR}$  where the bar pattern
speed and the angular velocity are equal, 
i.e. $\Omega_{p}=\Omega\,(R_{\rm CR})$.
We find that the corotation radius is at a distance $R_{\rm CR} = 33''$ 
($\sim5$ kpc) from the centre. This is consistent with Fig.~2 where the bar
seems to finish between $20''$ and $25''$ corresponding to about $28''-35''$ 
if deprojected onto the galaxy plane.

No outer Lindblad resonance (OLR) is found in the radial range of the observe
gas kinematics
because the condition $\Omega_{p}=\Omega(R)+k(R)/2$ is not satified for any 
value of the radius $R$.

\begin{figure}[h]
\resizebox{\hsize}{!}{\includegraphics{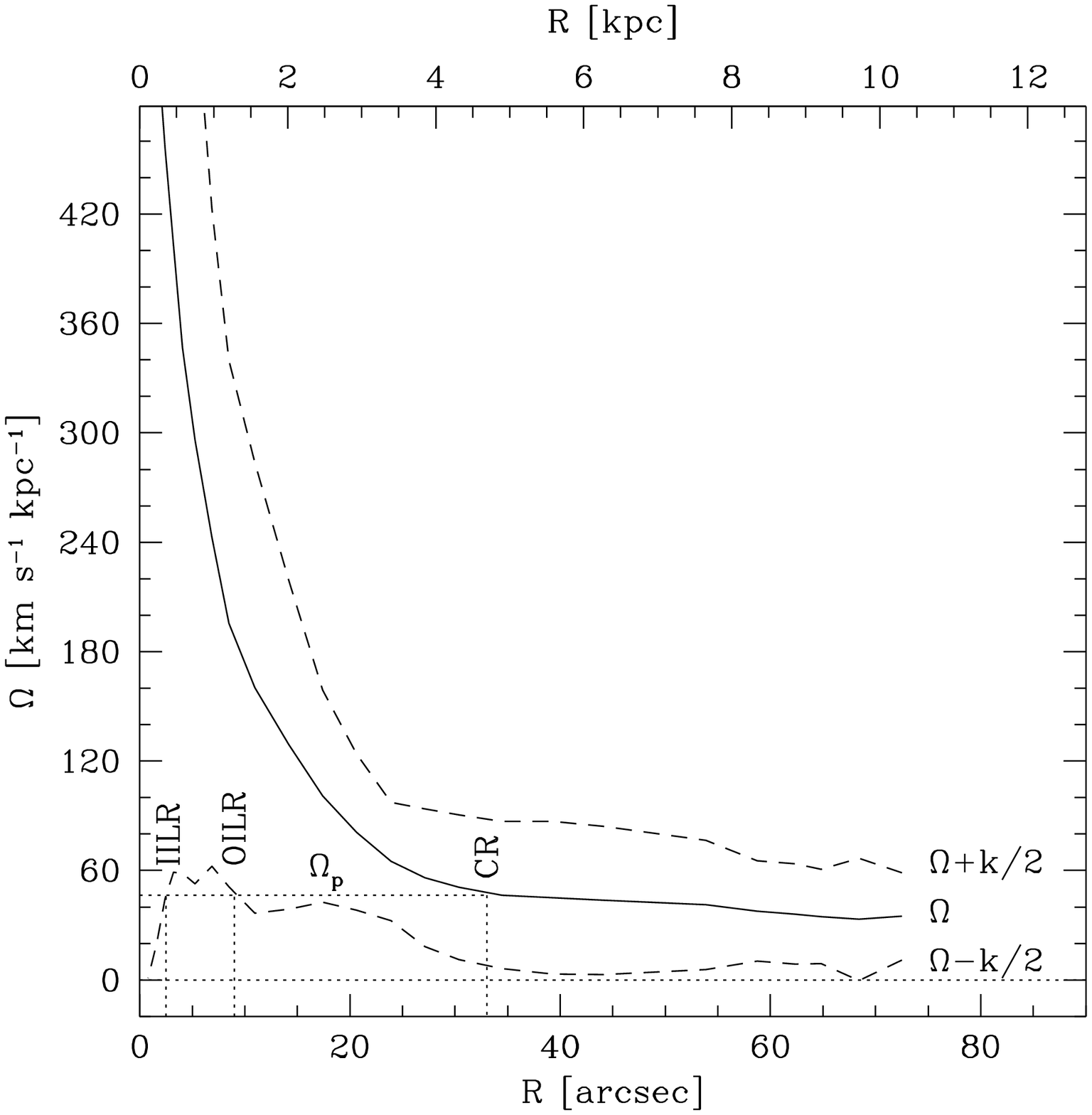}}
\caption{The angular velocity curve $\Omega$ and the derived curves
$\Omega\pm k/2$ in NGC~6221. The values of the angular velocity 
$\Omega$ and the epicyclic frequency $k$ are calculated from the 
model circular velocity deprojected on the galaxy plane. 
The horizontal line indicates the bar pattern speed $\Omega_p=46$
\kms\ kpc$^{-1}$. The positions of the two inner Lindblad resonances
($R_{\rm IILR}=3$\arcsec\ and $R_{\rm OILR}=9$\arcsec ) and
of the corotation ($R_{\rm CR}=33$\arcsec ) are showed. No
OLR has been found.  }
\end{figure}

\section{Discussion and Conclusion}

The possible link between bar component and nuclear activity has a long 
history either in the observational and in the theoretical field (e.g. 
Shlosman 1994). Recent observations have lead Ho et al. (1997) to disagree 
with this idea and to conclude that, even though the presence of a bar can 
enhance star formation in the bulge region, it does not affect significantly 
the active nucleus. Our data for NGC~6221 support this interpretation, 
because the physical properties of the central line emitting region indicate 
clearly that ionization comes from the presence of stellar sources. 
An AGN and its observational signatures might be obscured by gas and dust, 
but this is, of course, speculation.

The asymmetries of the velocity curves outside the bar region may be due to a 
tidal distorsion caused by NGC~6215. Koribalski (1996a) discovered 
an \hi\ bridge between NGC~6221 and NGC~6215 and later (Koribalski 1996b) 
suggested a possible interaction of NGC~6221 with the two newly discovered 
low-surface brightness galaxies nearby, namely 
\object{BK1} and \object{BK2}.

Elmegreen \& Elmegreen (1989) and later Keel (1996) proposed a dichotomy of 
bar properties between early-type (Sbc and earlier) and late-type galaxies 
(Sc and later) based on observational properties such as bar size, density 
profile of the bar, star formation, and so on. Early-type bars tend to be a 
dominant component in the galaxy with a flat density profile extending out to
the corotation radius, where they drive symmetric spiral arms (Combes \&  
Elmegreen 1993). On the contrary, bars in late-type galaxies are small with
exponentially decreasing density profiles, which extend only to the 
ILR radius. These kind of bars are unable to influence the wave pattern of the
stellar disk. This view is also supported by the model of Combes \& Elmegreen
(1993).

NGC~6221 is found to exhibit intermediate properties between early-type and
late-type barred spiral galaxies, as discussed below:
\begin{description}
\item [{\it i) Late-type features:\/}]  
The surface brightness profile shows  an exponential decrease; 
the star formation, i.e. the \ha\ emission, is detected along all the bar; 
the arm structure is typical of a late-type. 
\item [{\it ii) Early-type features:\/}] 
The edge of bar is located between the outer ILR radius and the corotation 
radius beyond the rising part of the velocity curve;
the strong gradient of gas velocity curve in the central $2$ kpc;
the dust-lane pattern is similar to that of the prototype SBb galaxies 
(see \object{NGC~1300} in Panels S8 and 154 in the CAG as example);
the radial profiles of the fluxes ratio of the \ha\ and \nii\ emission lines
reveal the presence of a ring structure associated with an ILR; moreover 
NGC~6221 is interacting with NGC~6215, and Elmegreen et 
al. (1990) have demonstrated the presence of bars in paired early-type spirals.
\end{description}
Indeed the case of NGC~6221 is analogous to that of certain other galaxies, 
notably NGC~1300 (Combes \&  Elmegreen 1993). It has been classified as a
late-type barred spiral because of the presence of gas, star formation 
and spiral arms, but it could be considered as an early-type due to
the dynamical properties of its bar.
 
Pfenniger (1992) has suggested that a major effect due to the presence of 
a bar could be that the galaxy evolves towards earlier morphological types.
Our study indicates that NGC~6221 is a typical case of a late-type barred 
spiral evolving to an earlier type. If our interpretation is correct, it is 
understandable that there is   so difficult to reach agreement about
its morphological classification which is Sbc(s) in RSA and SBc(s)
in RC3. Sometimes the criteria of classification in the Hubble sequence cannot
describe exactly the  complicated scenario of the evolution of galaxies,
as appears to be the case for NGC~6221.

\begin{acknowledgements} The DENIS team and in particular the operations team
at La Silla is warmly thanked for making this work possible.
The DENIS project is supported by the SCIENCE and the Human Capital and
Mobility plans of the European Commission under grants CT920791 and
CT940627, the European Southern Observatory, in France by the Institut
National des Sciences de l' Univers, the Education Ministery and the
Centre National de la Recherche Scientifique, in Germany by the State
of Baden-Wuerttenberg, in Spain by the DGICYT, in Italy by the
Consiglio Nazionale delle Ricerche, in Austria by the Science Fund
(P8700-PHY,P10036-PHY) and Federal Ministry of Science, Transport and
the Arts, in Brazil by the Foundation for the development of Scientific
Research of the State of Sao Paulo (FADESP). 
JCV acknowledges the support by a grant of the 
Telescopio Nazionale Galileo and Osservatorio Astronomico di Padova. 
WWZ acknowledges the support of the Jubil\"aums\-fonds der
\"Oster\-reichi\-schen National\-bank (grant 6323).
The research of MS is supported by the Austrian Science Fund projects 
P9638-AST and S7308.  
EMC acknowledges the head of IAC for hospitality during the preparation
of this paper.
This work was partially supported by grant PB94-1107 of the Spanish DGICYT.
The authors thank S. Garc\'{\i}a Burrillo for his valuable comments about
bars, and B. Koribalski for her useful information about  
NGC~6221 interaction with its companions.

\end{acknowledgements}


\begin{thebibliography}{}
\bibitem[1990]{bender} Bender, R. 1990, A\&A, 229, 441 

\bibitem[1994]{bender} Bender, R., Saglia, R.P., Gerhard. O.E.  1994, 
        MNRAS, 269, 785

\bibitem[1995]{bertola} Bertola, F., Cinzano, P., Corsini, E.M., Rix, H.-W., 
        Zeilinger, W.W. 1995, ApJ 448, L13

\bibitem[1996]{bertola} Bertola, F., Cinzano, P., Corsini, E.M., Pizzella, A., 
        Persic, M., Salucci, P. 1996, ApJ, 458, L67

\bibitem[1997]{bettoni} Bettoni, D., Galletta, G. 1997, A\&AS, 124, 61

\bibitem[1960]{brandt} Brandt, J.C. 1960, ApJ, 131, 293

\bibitem[1993]{combes} Combes, F., Elmegreen, B.G. 1993, A\&A, 271, 391

\bibitem[1991]{devacouleurs} de Vaucouleurs, G., de Vaucouleurs, A., Corwin, 
        H.G.Jr., Buta, R.J., Paturel, G., Fouqu\`e, P. 1991, Third Reference 
        Catalogue of Bright Galaxies, Springer-Verlag, New York (RC3) 

\bibitem[1997]{dottori} Dottori, H., Duval M.F., Carranza, G., Goldes,
  	G., Diaz, R., Paolantonio, S. 1996, RMxAA, 4, 136

\bibitem[1987]{durret} Durret, F., Bergeron, J. 1987, A\&A, 173, 219

\bibitem[1985]{elmegreen} Elmegreen, B.G., Elmegreen, D.M. 1985, ApJ, 288, 438

\bibitem[1989]{elmegreen} Elmegreen, B.G., Elmegreen, D.M. 1989, ApJ, 342, 677

\bibitem[1990]{elmegreen} Elmegreen, D.M., Bellin, A.D., Elmegreen, B.G. 
        1990, ApJ, 364, 415

\bibitem[1997]{epchtein} Epchtein, N., Debatz, B., Capoani, L., et al. 
        (49 authors) 1997, The Messenger, 87, 27

\bibitem[1997]{ho} Ho, L.C., Filippenko, A.V., Sargent, W.L.W. 1997, ApJ,
        487, 591

\bibitem[1996]{keel} Keel, W.C. 1996, In: Barred Galaxies, IAU Coll. 117,
        Buta R., Crocker D.A., Elmegreen B.G. (eds.), ASP Conf. Ser. 91, 
        ASP, San Francisco, p. 56

\bibitem[1996]{koribalski} Koribalski, B. 1996a, In: Barred Galaxies, 
        IAU Coll. 117, Buta R., Crocker D.A., Elmegreen B.G. (eds.),
        ASP Conf. Ser. 91, ASP, San Francisco, p. 172
 
\bibitem[1997]{koribalski} Koribalski, B., 1996b, In: The Minnesota
        Lectures on Extragalactic Neutral Hydrogen, Skillman E.D. (ed.),
	ASP Conf. Series 106, ASP, San Francisco, p. 238

\bibitem[1984]{pence} Pence, W.D., Blackman, C.P. 1984, MNRAS 207, 9

\bibitem[1992]{pfenniger} Pfenniger, D. 1992, In: Physics of Nearby 
        Galaxies: Nature or Nurture?, 27th Rencontre de Moriond,   
        Thuan T.X., Balkowsky C., Tran Thanh Van J. (eds.), Editons 
        Frontieres, Gif-sur-Yvette, p. 519

\bibitem[1979]{philipps} Philipps, M.M. 1979, ApJ 227, L121

\bibitem[1979]{roberts} Roberts, W.W.Jr., Huntley, J.M., van Albada, G.D. 
        1979, ApJ, 233, 67

\bibitem[1994]{sandage} Sandage, A., Bedke, J. 1994, The Carnegie Atlas of 
        Galaxies, Carnegie Institution, Flintridge Foundation, Washington (CAG)

\bibitem[1981]{sandage} Sandage, A., Tammann, J. 1981, A Revised Shapley-Ames
	Catalog of Bright Galaxies, Carnegie Institution, Washington (RSA)
 
\bibitem[1994]{shlosman} Shlosman, I., 1994, Mass-Transfer Induced Activity 
        in Galaxies, Cambridge University Press, Cambridge 

\bibitem[1996]{zeilinger} Zeilinger, W.W., Pizzella, A., Amico, P., Bertin, 
        G., Bertola, F., Buson, L.M., Danziger, I.J., Dejonghe, H., Sadler,
        E.M., Saglia, R.P., De Zeeuw, P.T. 1996, A\&AS 120, 257

\end{thebibliography}
\end{document}